\documentclass[letter,nobibnotes,nofootinbib]{revtex4}

\usepackage{amsmath,amssymb}
\usepackage{epsfig}
\usepackage{graphicx}
\usepackage{xspace}

\newcommand{\be}{\begin{equation}}
\newcommand{\ee}{\end{equation}}
\newcommand{\bL}{\begin{Large}}
\newcommand{\eL}{\end{Large}}
\newcommand{\ba}{\begin{eqnarray}}
\newcommand{\ea}{\end{eqnarray}}
\newcommand{\bc}{\begin{center}}
\newcommand{\ec}{\end{center}}
\newcommand{\bfig}{\begin{figure}}
\newcommand{\efig}{\end{figure}}

\begin{document}

\title{Scaling properties of elastic proton-proton scattering at LHC energies}

\author{C. Baldenegro}%
 \email{c.baldenegro@cern.ch}
\affiliation{%
\'{E}cole Polytechnique, Laboratoire Leprince-Ringuet, Av. Chasles, 91120 Palaiseau, France\\
}%

\author{C. Royon}%
 \email{christophe.royon@cern.ch}
\affiliation{%
 Department of Physics and Astronomy, The University of Kansas, Lawrence, KS 66045, USA \\
}%

\author{A.M. Stasto}%
 \email{ams52@psu.edu}
\affiliation{%
 Department of Physics, Penn State University, University Park, PA 16802, USA\\
}%


\date{\today}

\begin{abstract}
The TOTEM Collaboration has measured the differential cross section of elastic proton-proton scattering $\mathrm{d}\sigma/\mathrm{d}|t|$ at $\sqrt{s} = 2.76$, 7, 8, and 13 TeV. It is observed that all $\mathrm{d}\sigma/\mathrm{d}|t|$ data points vary in the same way as a function of the center-of-mass energy in the so-called ``dip'' and ``bump'' regions. These features hint at possible universal properties of elastic scattering. Based on these empirical observations, and taking inspiration from saturation models, we propose a simple scaling law for proton-proton elastic scattering at LHC energies. 
 We find that the $\mathrm{d}\sigma/\mathrm{d}|t|$ at LHC energies fall onto a universal curve when they are mapped to the scaling variables $  \mathrm{d}\sigma/\mathrm{d}|t| \times (s/\text{TeV}^2)^{-0.305} $ versus $(s/\text{TeV}^2)^{0.065} (|t|/\text{GeV}^2)^{0.72}$. Some implications of this scaling law in the impact parameter picture of the scattering amplitudes are explored.

\end{abstract}

\maketitle

\section{Introduction}

In this Letter, we introduce a scaling law based on empirical observations,  
describing the recent measurements~\cite{TOTEM:2011vxg,TOTEM:2015oop,TOTEM:2018psk,TOTEM:2017sdy} by the TOTEM Collaboration of the elastic differential cross sections $d \sigma/d|t|$ as a function of $|t|$, the magnitude of the four-momentum transfer squared.

High-energy elastic scattering of hadrons $h_1 h_2 \to h_1 h_2$ is described in terms of $t-$channel exchanges of colorless states that do not carry any flavor quantum numbers. At lower energies, such as at the Intersecting Storage Rings (ISR) at CERN, elastic scattering is due to pomeron, reggeon, and meson exchanges, the latter two corresponding to a quark-antiquark pair at the lowest order in the perturbation theory. At very high energies, the exchanges involving compound structures of gluons, namely pomeron and odderon, become dominant (the pomeron is dominant at high energies, the odderon intercept is predicted to be smaller, thus giving a subleading contribution). In a recent publication, the D0 and TOTEM Collaborations discovered the $C$-odd exchange, the odderon exchange, by comparing $pp$~\cite{TOTEM:2011vxg,TOTEM:2015oop,TOTEM:2018psk,TOTEM:2017sdy} and $p \bar{p}$~\cite{D0:2012erd} interactions~\cite{TOTEM:2020zzr}.

Elastic scattering of hadrons has been measured since the 1950s at different center-of-mass energies ($pp$ and $p \bar{p}$ interactions), for instance at the ISR and at the SPS at CERN~\cite{Erhan:1984mv,Breakstone:1985pe,UA4:1985oqn,UA4:1986cgb,Nagy:1978iw}
and more recently at the Large Hadron Collider (LHC) and in cosmic 
ray experiments~\cite{Zyla:2020zbs}.  The $pp$ and $p \bar{p}$ total cross sections show an increase with $s$. The $d \sigma/d|t|$ elastic cross sections in $pp$ collisions as a function of $|t|$ have some maxima and minima, which we call ``bump'' and ``dip'' hereafter, whereas this was not observed for $p \bar{p}$ at high energies~\cite{TOTEM:2020zzr}.

In this Letter, we focus on elastic scattering at LHC energies, where elastic scattering is dominated by pomeron exchange. This can be seen in Ref.~\cite{TOTEM:2020zzr}, where the value of the elastic cross section at the bump divided by the value at the dip, $R$, show a flat dependence above 1 TeV. Scaling properties at lower $\sqrt{s}$ (ISR energies) have been investigated previously, see \cite{DiasDeDeus:1973lde, Dremin:2012qd, Csorgo}.

The TOTEM experiment recently published the $pp$ elastic scattering cross section $d \sigma/d|t|$ at the LHC at center-of-mass energies of 2.76, 7, 8 and 13 TeV~\cite{TOTEM:2011vxg,TOTEM:2015oop,TOTEM:2018psk,TOTEM:2017sdy}. In order to extrapolate the TOTEM $pp$ measurements to the Tevatron center-of-mass energy of 1.96 TeV, a set of eight reference points (such as the dip and bump) were identified~\cite{TOTEM:2020zzr} in the common kinematical domain in $|t|$ where D0 and TOTEM have data. These reference points are characteristic of the shape of elastic $\mathrm{d}\sigma/\mathrm{d}|t|$ at high energies.
The TOTEM elastic scattering $\mathrm{d}\sigma/\mathrm{d}|t|$ measurements showed the feature that all $|t|$ and $d \sigma/d|t|$ values at the reference points show the same dependence as a function of $\sqrt{s}$. This means that the elastic scattering differential cross sections as a function of $|t|$ show a translation in the ($|t|$,$d \sigma/d|t|$) plane as also shown in Fig.~\ref{fig:data}. All points on the curves vary in the same way as a function of $\sqrt{s}$,  suggesting scaling properties of the differential cross section as a function of $s$.

The structure of this Letter is the following.
In Section \ref{sec:section2}, we describe the scaling that we observe in data. In Section \ref{sec:section3}, we interpret this scaling in impact parameter space, before presenting a conclusion and an outlook  in Section \ref{sec:section4}.

\section{Scaling properties of elastic scattering}\label{sec:section2}
\begin{figure*}
\centering

\includegraphics[width = 0.47\textwidth]{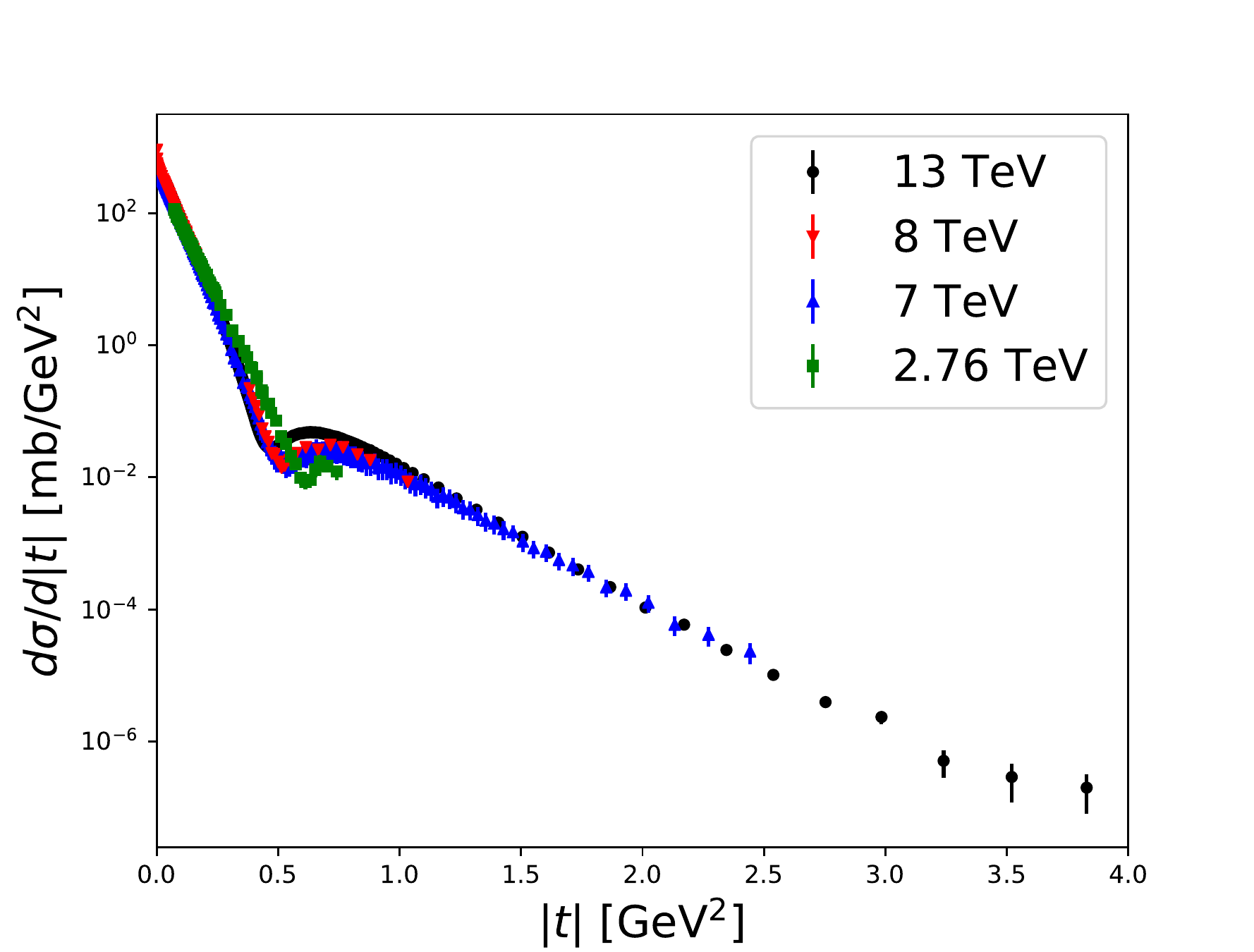}
\includegraphics[width = 0.49\textwidth]{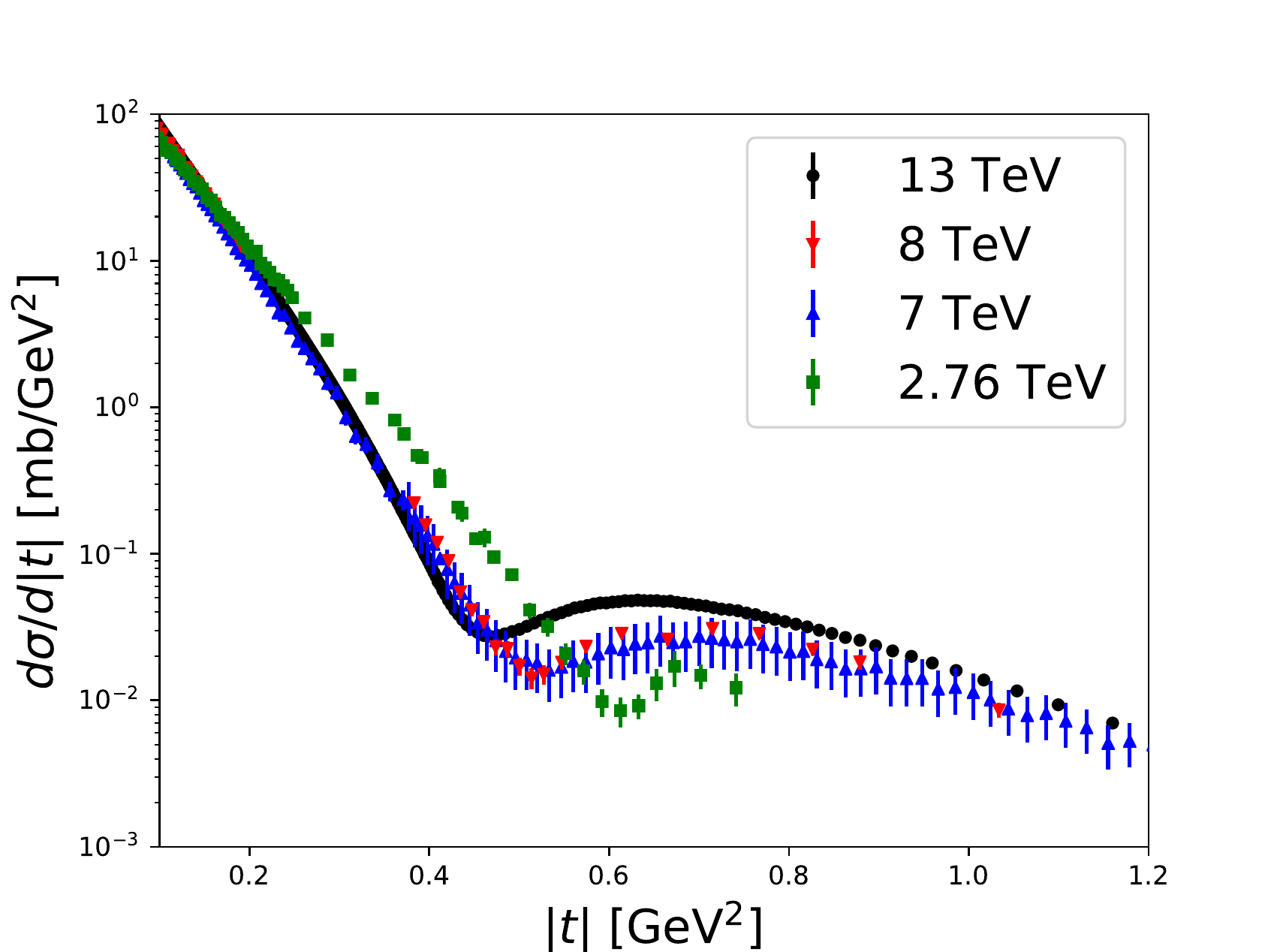}
\caption{ 
 Left: Elastic proton-proton differential cross section $\mathrm{d}\sigma/\mathrm{d}|t|$ at $\sqrt{s} = $2.76, 7, 8, and 13 TeV measured by the TOTEM Collaboration~\cite{TOTEM:2011vxg,TOTEM:2015oop,TOTEM:2018psk,TOTEM:2017sdy}.
 Right: same data zoomed into the dip and bump region with $0.1 < |t| < 1.2$ GeV$^2$ for visualization purposes.
}
\label{fig:data}
\end{figure*}

As evident from Fig.~\ref{fig:data} and from Fig.~1 of Ref.~\cite{TOTEM:2020zzr}, the 
 TOTEM elastic cross section measurements as a function of $|t|$ for different $\sqrt{s}$ values do not vary in an independent way. They exhibit a pattern, where the curves are shifted, and this hints to the presence of a scaling variable in data.
 
 \subsection{The quality factor method}
 
Given a set of data points $(x_i,y_i)$, we want to know whether the $y_i$ values  can be parametrized as a smooth and continuous function of the variable 
in the abscissa $x_i$. Since the function $y_i = f(x_i)$ that describes the data is not known a priori, we calculate a quantity known as the quality factor (QF). The QF is indicative of the degree to which the set of data points $(x_i,y_i)$ can be parametrized as a function of a smooth, continuous function while being agnostic to the underlying shape of such a functional form, if it exists. The QF is defined in a way such that it quantitatively describes how close consecutive data points vary between each other. This method has been used to test scaling properties of the structure function extracted from deep-inelastic scattering data \cite{Gelis:2006bs, Beuf:2008mf, Marquet:2006jb, Royon:2008jv}.

In our case, the $x_i$ and $y_i$ are the variables that show a potential scaling behavior and are respectively a function of $|t_i|$ and elastic $(d \sigma/d|t|)_i$. To obtain variables that vary smoothly over the full range of the data, we first take the logarithm of these variables, $u_i=\log(x_i)$ and $v_i=\log(y_i)$ (both $\mathrm{d}\sigma/\mathrm{d}|t|$ and $|t|$ span several orders of magnitude in units of mb GeV$^{-1}$ and GeV$^2$, respectively). Then, $u_i$ and $v_i$ are shifted such that the minimum values of $u_i$ and $v_i$ are 0, and then they are rescaled such that their maximum values are 1. The QF is defined as,

\begin{eqnarray}
{\rm QF}= \left[ \Sigma_i \frac{(v_{i+1}-v_i)^2 \times \Delta v_{i+1} \times \Delta v_i}{(u_{i+1}-u_i)^2 + \epsilon^2} \right] \; ,
\end{eqnarray}

where the $\Delta v_i$ are the uncertainties on $v_i$ and $\epsilon$ is a small constant to regularize divergences when $u_{i+1}=u_i$, with $\epsilon^2 = 10^{-7}$. The value of $\epsilon^2$ is much smaller than the smallest difference between consecutive points. The value of QF is large when  $v_i$ and $v_{i+1}$ differ by a large amount while the $u_i$ and $u_{i+1}$ are close to one another, or, in other words, when there is no smooth dependence of $v_i$ as a function of $u_i$. The QF is then minimized as a function of the scaling constants. Unfortunately, it is not straightforward to extract an uncertainty on the scaling constants fit with this method (this is not a $\chi^2$ scan). However, we verify the stability of the results with variations of the scaling constants.

 \subsection{Understanding scaling in elastic scattering data at high energies}

The first step of the scaling study is to find a new variable, which we call $t^{**}$, for which $(s/1 \text{TeV}^2)^{-\alpha} d \sigma/d|t|$ as a function of $t^{**}$ does no longer depend on $\sqrt{s}$, where $\alpha$ is a constant to be fit to the data with the QF method. 

We start by assuming the following expression for $t^{**} \equiv t^*/s^B = (s/|t|)^A |t| / s ^B$, where $A$ and $B$ are scaling constants to be fit to the data. In this Letter, we work with units of TeV$^2$ and GeV$^2$ for $s$ and $t$, respectively, which renders the scaling variable $t^{**}$ to be of order 1.
The choice of this functional form for the variable $t^*$ is motivated by the geometrical scaling in the context of saturation models. In particular, the form of the variable $t^*=|t|(s/|t|)^A$ is similar to the  scaling variable with saturation scale, used for example in  \cite{McLerran:2014apa,Praszalowicz:2015dta}  in the context of the $p_\mathrm{T}$ spectra of particle multiplicities at the LHC. Empirically, we find that an additional power of $s$ is needed to obtain the desired scaling in the data. This is needed to describe the position of the dip and the bump as a function of $s$.

\begin{figure*}
\centering
\includegraphics[width=0.49\textwidth]{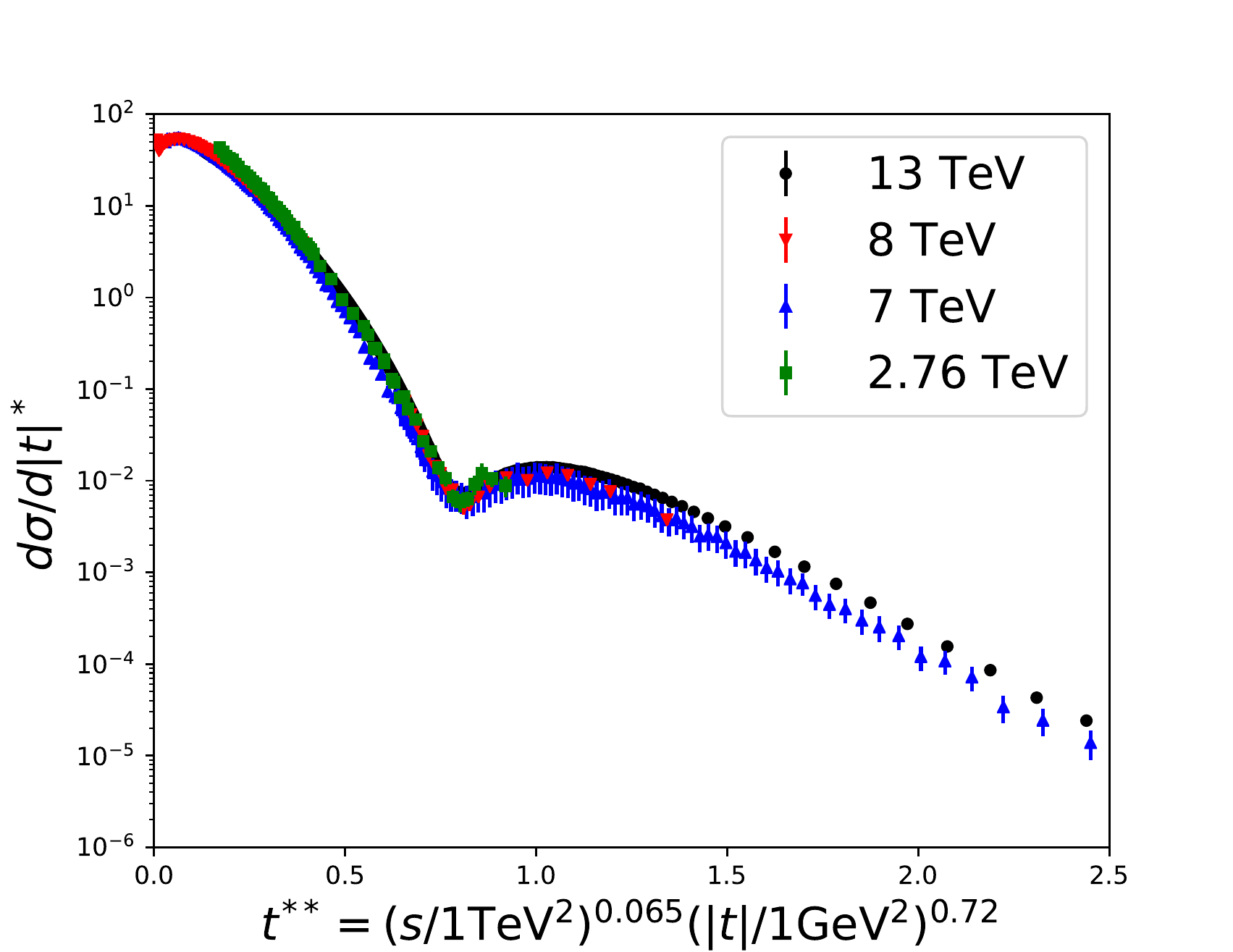}
\includegraphics[width=0.49\textwidth]{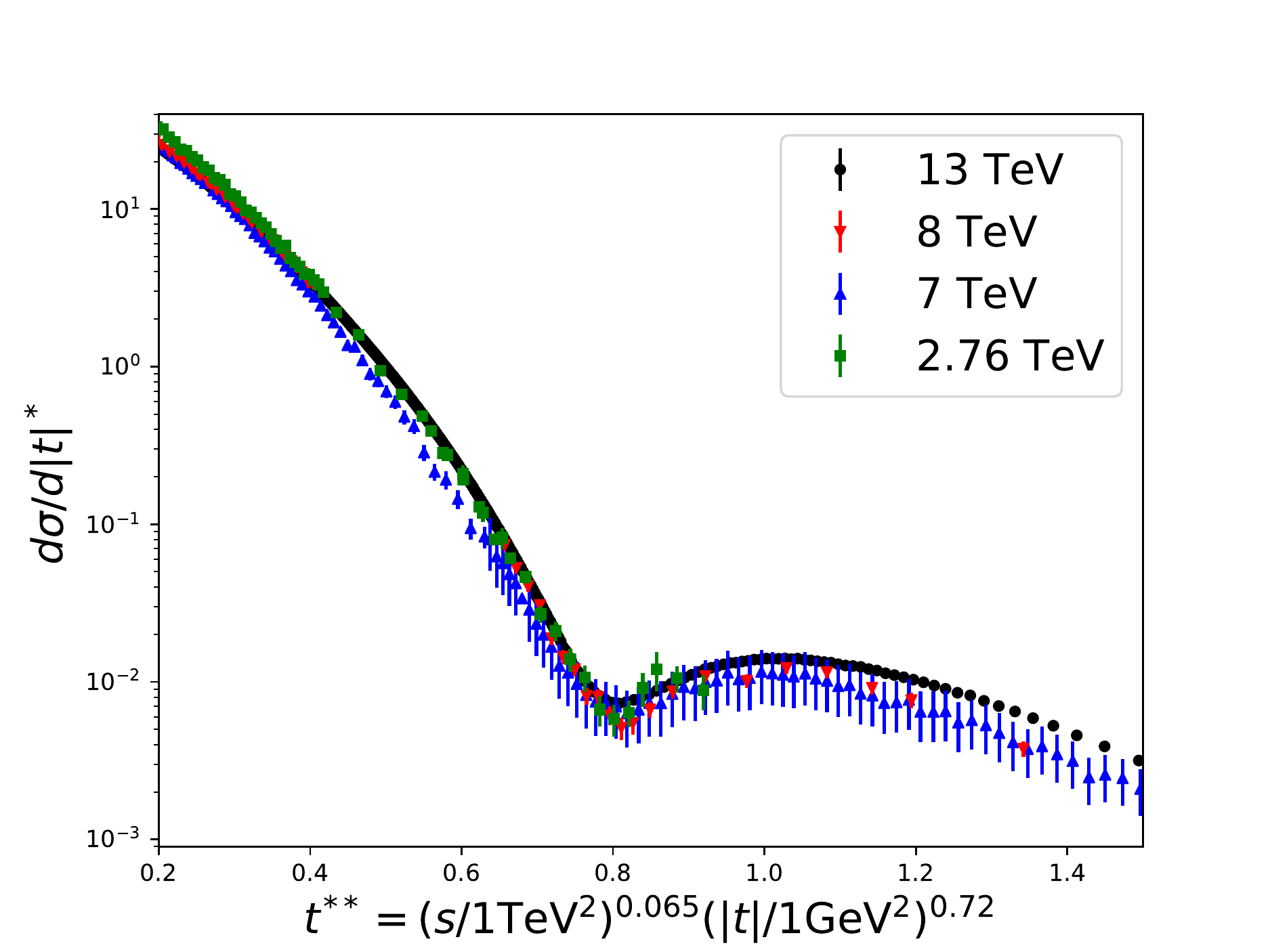}
\caption{ 
Left: $\mathrm{d}\sigma/\mathrm{d}t^*$ as a function of $t^{**}$ showing the scaling of all TOTEM elastic scattering data at $\sqrt{s}=$ 2.76, 7, 8 and 13 TeV. Right: same results, but zoomed into the dip and bump region for visualization purposes.
}
\label{fig2}
\end{figure*}

We use the QF method described in the previous section to fit $A$ and $B$. Our first observation is that the elastic scattering $\mathrm{d}\sigma/\mathrm{d}|t|$ cross sections measured by the TOTEM Collaboration at 2.76, 7, 8 and 13 TeV scale if we display them as $\mathrm{d}\sigma/\mathrm{d}|t|^*$ as a function of $t^{**}$ as shown in Fig.~\ref{fig2}. Looking for a minimum in the QF value, we find there is a set of $(A,B)$ pairs that lie on a valley in parameter space. The pairs of $(A,B)$ values that yield a minimum QF are strongly correlated, as shown on the left panel of Fig.~\ref{valley}. A linear fit to this correlation plot leads to $ B = (1.000 \pm 0.003)A - 0.065\pm 0.001$. We thus have a single parameter to be fit to the data, $t^{**}=(s/\mathrm{TeV}^2)^{0.065} (|t|/\mathrm{GeV}^2)^{1-A}$. This means that $t^{**}$ has a universal behavior as a function of $s$ and this dependence is rather soft as $s^{0.065}$. Conversely, $t$ has a universal behavior as a function of $s$. In particular, the positions of the dip and the bump are fixed in $t^{**}$ and do not longer depend on $s$.  This factor of $s^{0.065}$ describes the dependence of the position in $|t|$ of the dip and the bump with $s$, which is known to have a soft dependence on $s$. 
 
\subsection{Results on scaling}

The last step is to find the universal behavior of $d \sigma/d|t|$ as a function of $s$, which originates from scaling. We start with the assumption that $\mathrm{d}\sigma/\mathrm{d}t^*$ scales perfectly as a function of $t^{**}$. Based on this, the goal is to find the constant $\alpha$ for which $(s/\mathrm{TeV}^2)^{-\alpha} \mathrm{d}\sigma/\mathrm{d}|t|$ as function of $t^{**}$ does not show any residual dependence on $s$.

To obtain a relation between $\alpha$ and $A$, the parameter defined above in $t^{**}$, we reformulate

 \begin{equation}
\frac{d\sigma}{dt^*} = \frac{d\sigma}{dt} \frac{dt}{dt^*} \; .
\end{equation}

the conversion factor $dt/dt^*$ as a function of $t^{**}$ leads to

\begin{equation}
\frac{dt}{dt^*} = \frac{1}{1-A} \frac{1}{(s/\mathrm{TeV}^2)^A} \big(t^{**} (s/\mathrm{TeV}^2)^{B-A}\big)^\frac{A}{1-A} \; ,
\end{equation}

assuming $B=A-0.065$ to reduce the analysis to a one-parameter fit, leads to

\begin{equation}
\frac{d\sigma}{dt^*} = \frac{d\sigma}{dt} \times s^{A \frac{A-1.065}{1-A}} \times f(t^{**}) = (s/\mathrm{TeV}^2)^{-\alpha} \frac{d\sigma}{dt}  f(t^{**}) \;  ,
\end{equation}

where $f(t^{**})$ is a function that carries the dependence on $t^{**}$ from the $dt/dt^*$ term. Then, on grounds that $\frac{d\sigma}{dt^*}$ should not depend on $s$ (i.e., assuming that scaling is perfect), it follows that the product $s^{-\alpha} \times \frac{d\sigma}{dt}$ must not depend on $s$, where $\alpha \equiv A \frac{1.065-A}{1-A} $. The implication here is that the scaling, which acts on both $|t|$ and $\mathrm{d}\sigma/\mathrm{d}|t|$, can be fully described by a single constant.

\begin{figure*}
\centering
\includegraphics[width=0.47\textwidth]{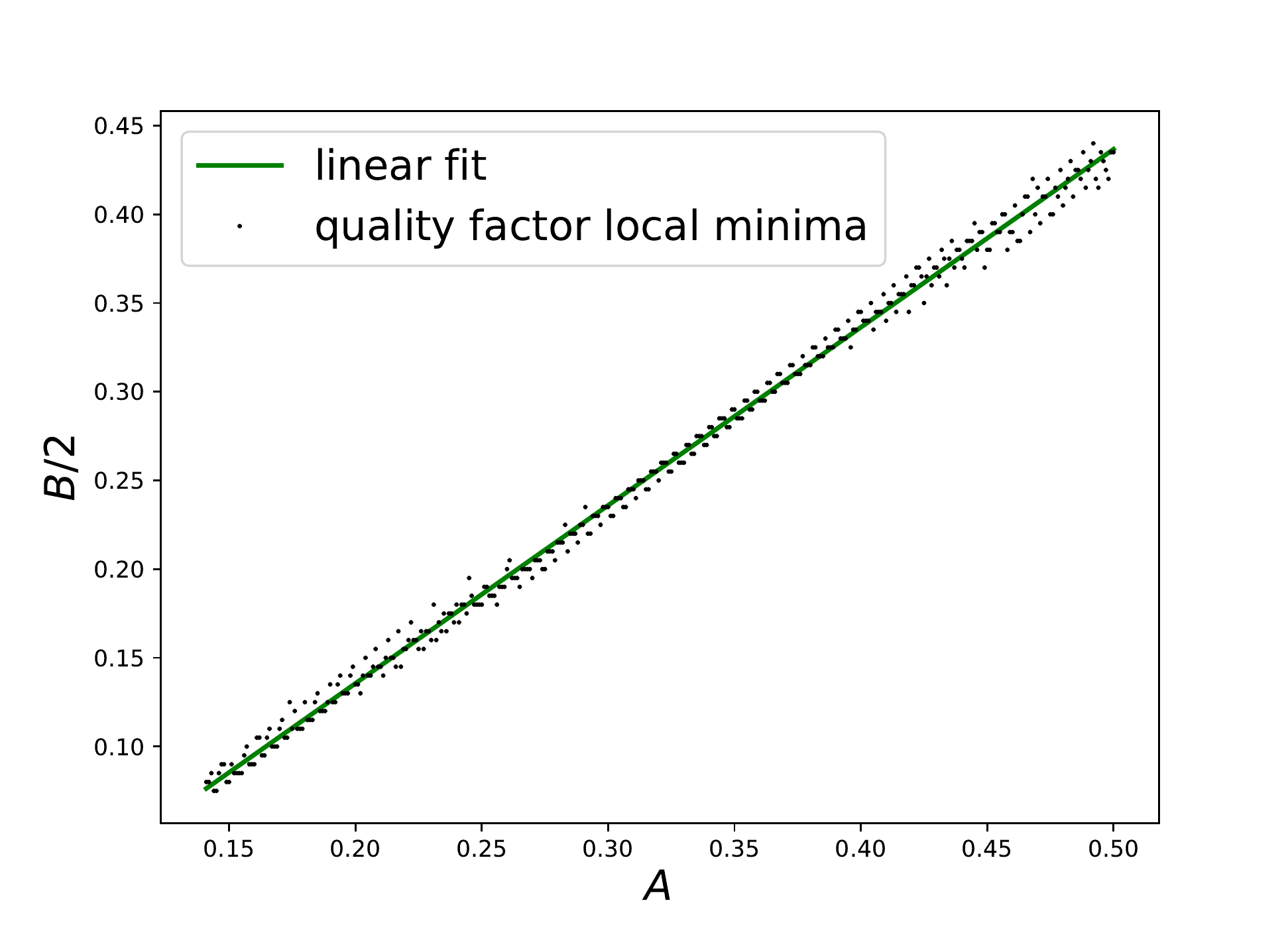}
\includegraphics[width=0.47\textwidth]{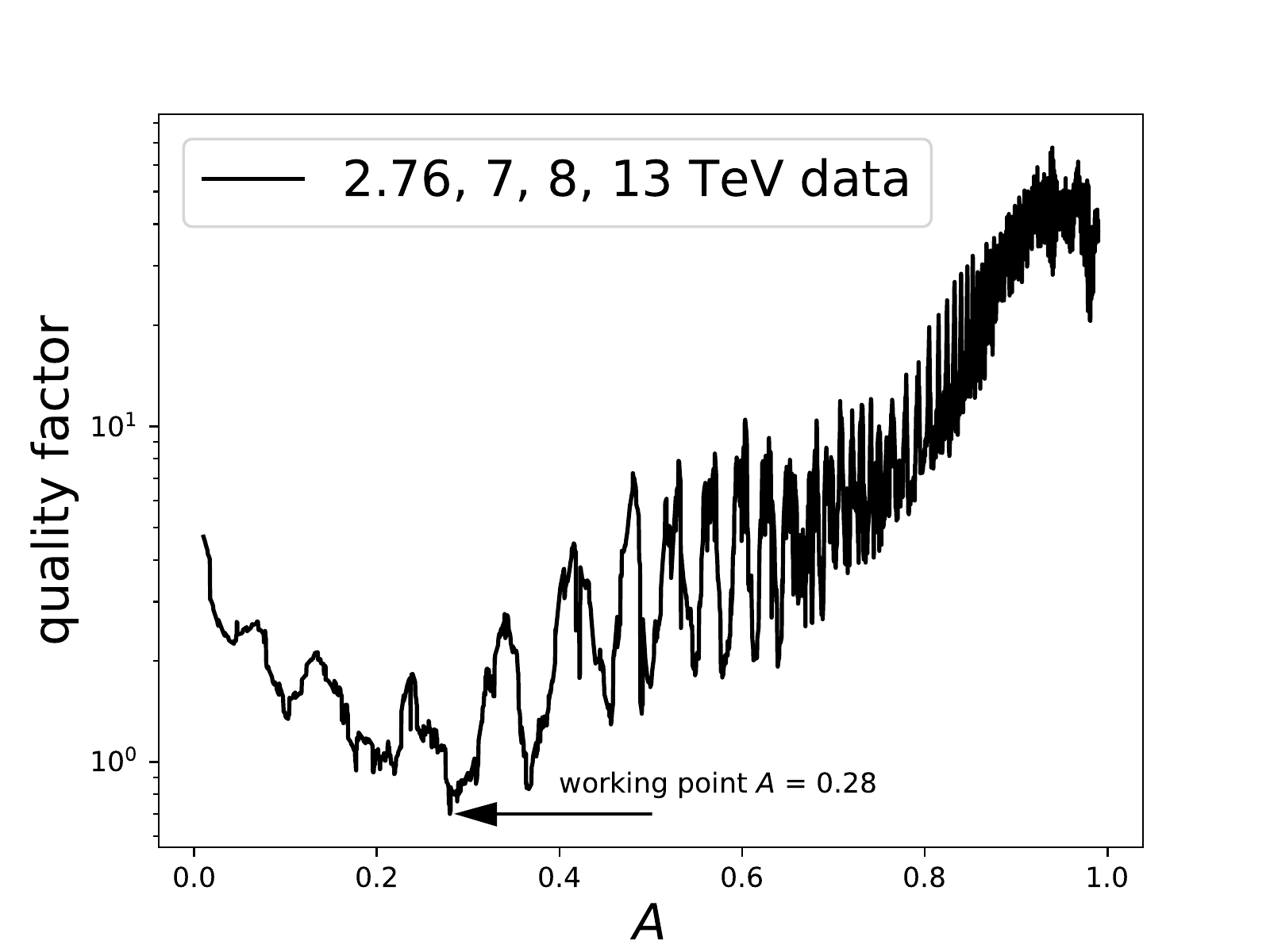}
\caption{ 
Left: Correlation between the $A$ and $B$ scaling constants that lead to a minimum QF. Right: One-dimensional scan of the QF value as a function of the $A$ parameter after assuming $B=A-0.065$. The working point $A = 0.28$ where the QF is minimized is used throughout this Letter. The oscillations on the QF value as a function of $A$ are typical of this fitting method.
}
\label{valley}
\end{figure*}

We fit the value of $A$ using all elastic proton-proton scattering data measured by TOTEM at 2.76, 7, 8, and 13 TeV~\cite{TOTEM:2011vxg,TOTEM:2015oop,TOTEM:2018psk,TOTEM:2017sdy}. To use the QF method described in the previous section, we choose $u_i$ and $v_i$ respectively as $\ln( t^{**})$ and $\ln [ (s/\mathrm{TeV}^2)^{-\alpha} d \sigma/d|t|]$ where $t^{**}=(s/\mathrm{TeV}^2)^{0.065} (|t|/\mathrm{GeV}^2)^{1-A}$ and  $ \alpha= A(1.065-A)/(1-A)$, since both $t^{**}$ and $\mathrm{d}\sigma/\mathrm{d}|t|$  span orders of magnitude. After normalizing the $u_i$ and $v_i$ between 0 and 1, we obtain a value of 
$A=0.28$  (which corresponds to $\alpha=0.305$) leading to a QF of 0.7 as shown on the right panel of Fig.~\ref{valley}. 

\begin{figure*}
\begin{center}
\includegraphics[width=0.49\textwidth]{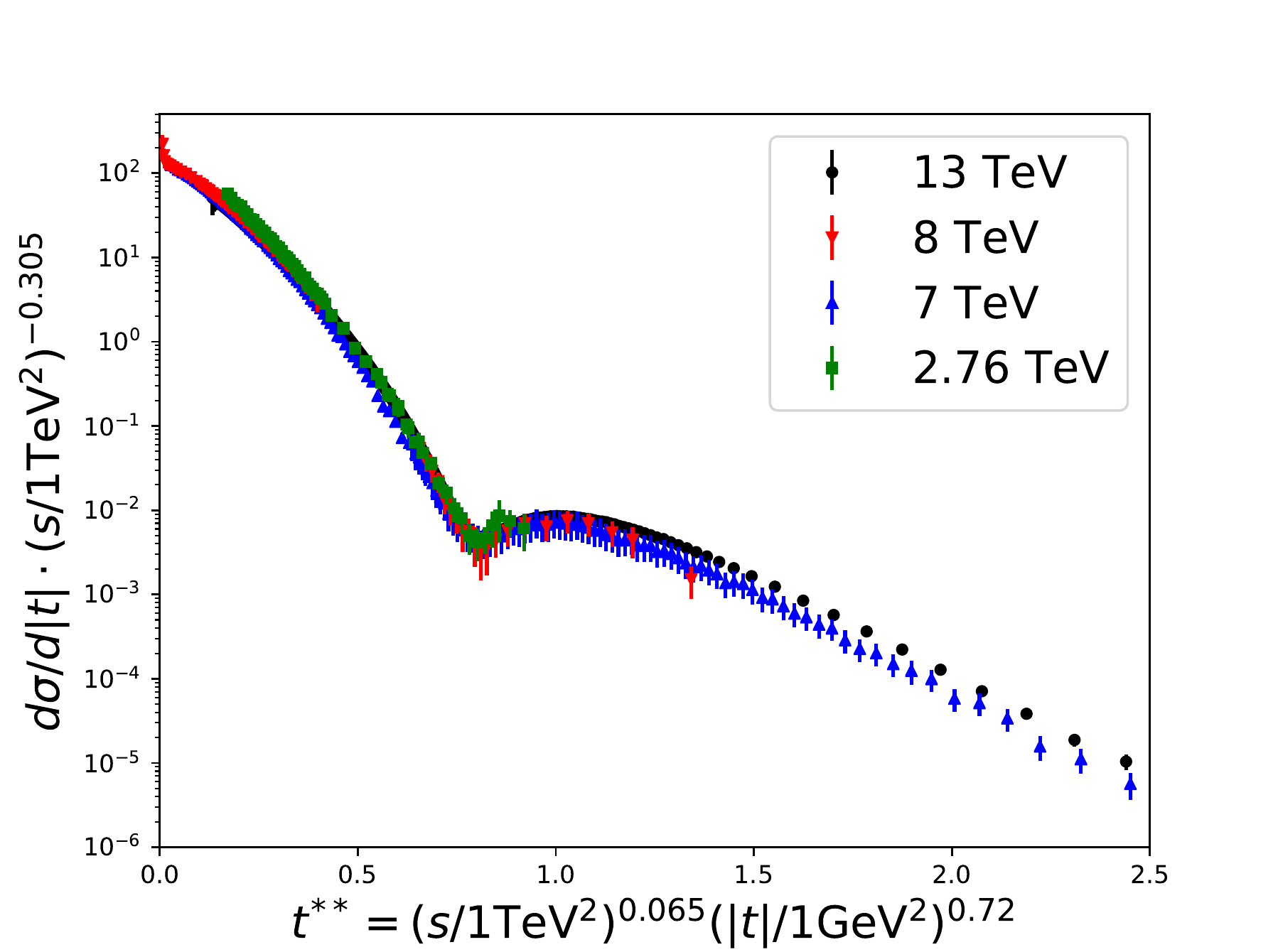}
\includegraphics[width=0.49\textwidth]{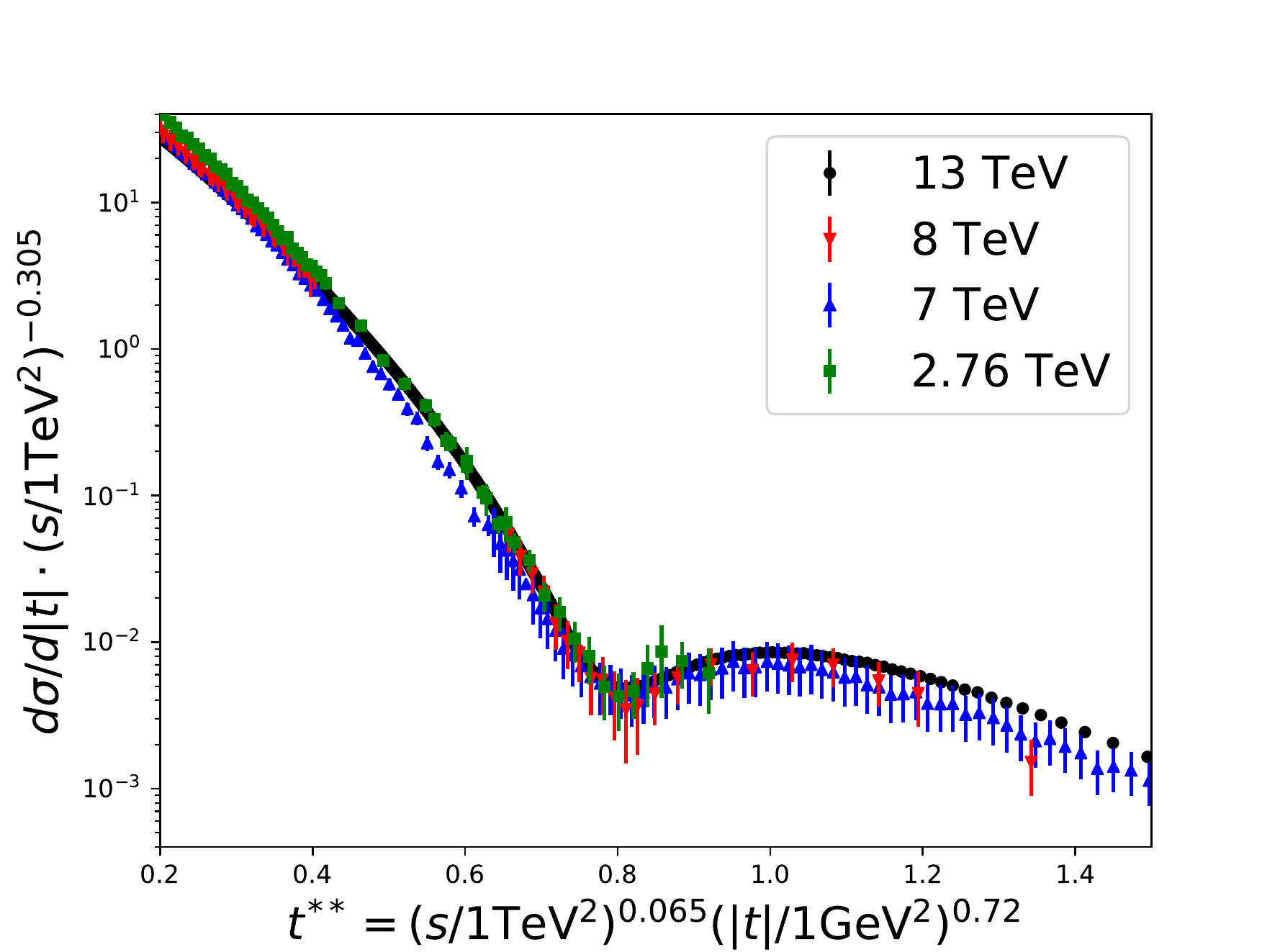}
\caption{Left: $(s/\mathrm{TeV}^2)^{-0.305} d \sigma/d|t|$ as a function of $t^{**}$ showing the scaling of all TOTEM $d \sigma/d|t|$ data in these variables. Right: same results, but zoomed into the dip and bump region for visualization purposes.} 
\end{center}
\label{scalingfinal}
\end{figure*}

The scaling result is shown in Fig.~\ref{scalingfinal} where we display $(s/\mathrm{TeV}^2)^{-0.305} d \sigma/d|t| $ as a function of $t^{**}=(s/ $1 TeV$^2)^{0.065} (|t|/$1 GeV$^2)^{0.72}$. The TOTEM data collapse onto a universal curve in a wide region of $t^{**}$, with the exception of very small and large values.

We have checked the stability of the scaling parameter. Table \ref{tab:my_label} displays the $A$ parameter values, as well as QF and the number of fit points using the QF method for different $|t|$ selections on data. We notice the very good stability of $A$ for all selections of $|t|$ ranges.

\begin{table}[]
    \centering
    \begin{tabular}{|c||c|c|c|}
    \hline
     data    & number of data points & $A$ & QF \\ \hline
    all     & 599 & 0.28 & 0.70 \\
    $|t| \ge 0.01$ GeV$^2$ & 557 & 0.28 & 0.70 \\
    $|t| \ge 0.05$  GeV$^2$ & 545 & 0.28 & 0.70 \\
    $|t| \ge 0.1$  GeV$^2$ & 476 & 0.28 & 0.69 \\
    $|t| \ge 0.2$  GeV$^2$& 351 & 0.28 & 0.65 \\
    $|t| \le 1.$  GeV$^2$& 548 & 0.28 & 0.57 \\
    $|t| \le 0.5$  GeV$^2$& 444 & 0.31 & 0.25 \\ \hline
    \end{tabular}
    \caption{Stability of the $A$ parameter selecting only parts of data at 2.76, 7, 8 and 13 TeV.}
    \label{tab:my_label}
\end{table}

\section{Implications in impact parameter space}\label{sec:section3}

\subsection{Fits to the TOTEM elastic scattering $\mathrm{d} \sigma/\mathrm{d}|t|$ data }

To explore possible physical implications of the scaling behavior observed in $s$ and $t$ space, we analyze the scattering amplitudes in impact parameter $b$. The $b$ dependent amplitudes can be obtained provided we have the parametrization of the scattering amplitudes in $|t|$. In order to explore this representation of the scattering amplitudes, the differential cross section $\mathrm{d}\sigma/\mathrm{d}|t|$ must be fit. We first obtain the $s$ dependence of $\mathrm{d}\sigma/\mathrm{d}|t|$ from the scaling described in the previous section, and we fit the $|t|$ dependence using a double exponential fitting formula, as described later in this section.

We calculate the real part of the profile function, $\mathrm{Re}(\Gamma(s,b))$, via a Fourier--Hankel transform of the elastic scattering amplitude $A (s, t)$,
\begin{eqnarray}
\mathrm{Re}(\Gamma(s,b)) = \frac{1}{4 \pi i s} \int_0^{\infty} dq \, q  \, J_0(qb) \,  A(s,t=-q^2) \; ,
\end{eqnarray}

where $J_0$ is the 0th order Bessel function. The total cross section can be calculated from Re$\Gamma$ as follows 

\begin{eqnarray}
\sigma_{\rm tot} &=& 2 \int d^2\mathbf{b} \, {\rm Re} \Gamma(s,b)  \; .
\end{eqnarray}

The differential elastic cross section reads
\begin{eqnarray}
\frac{\mathrm{d}\sigma}{\mathrm{d}|t|}= \frac{1}{16 \pi s^2 } |A(s,t)|^2 \; .
\label{sigmael}
\end{eqnarray}

We also define the rescaled amplitude that is convenient for fitting
\begin{eqnarray}
\frac{\mathrm{d} \sigma}{\mathrm{d}|t|} = |\mathcal{A}(s,t)|^2 \; ,
\end{eqnarray}
such that
\begin{eqnarray}
A(s,t) = 4 \sqrt{\pi} s \mathcal{A}(s,t) \; .
\end{eqnarray}
It leads to
\begin{eqnarray}
{\rm Re} \Gamma(s,b)= {\rm Re} \frac{1}{\sqrt{\pi} i} \int_0^{\infty} dq q J_0(qb) \mathcal{A}(s, t=-q^2)\, .
\end{eqnarray}

The amplitude $\mathcal{A}(s, t=-q^2)$ must be fit as a function of $|t|$ with the $s$ dependence given by the scaling observed in the previous section.

\subsection{Fits to the TOTEM data}

In this section, we describe the fits of $\mathrm{d}\sigma/\mathrm{d}|t|$ as a function of $|t|$. The $s$ dependence is given by scaling.

\begin{figure*}
\centering
\includegraphics[width=0.47\textwidth]{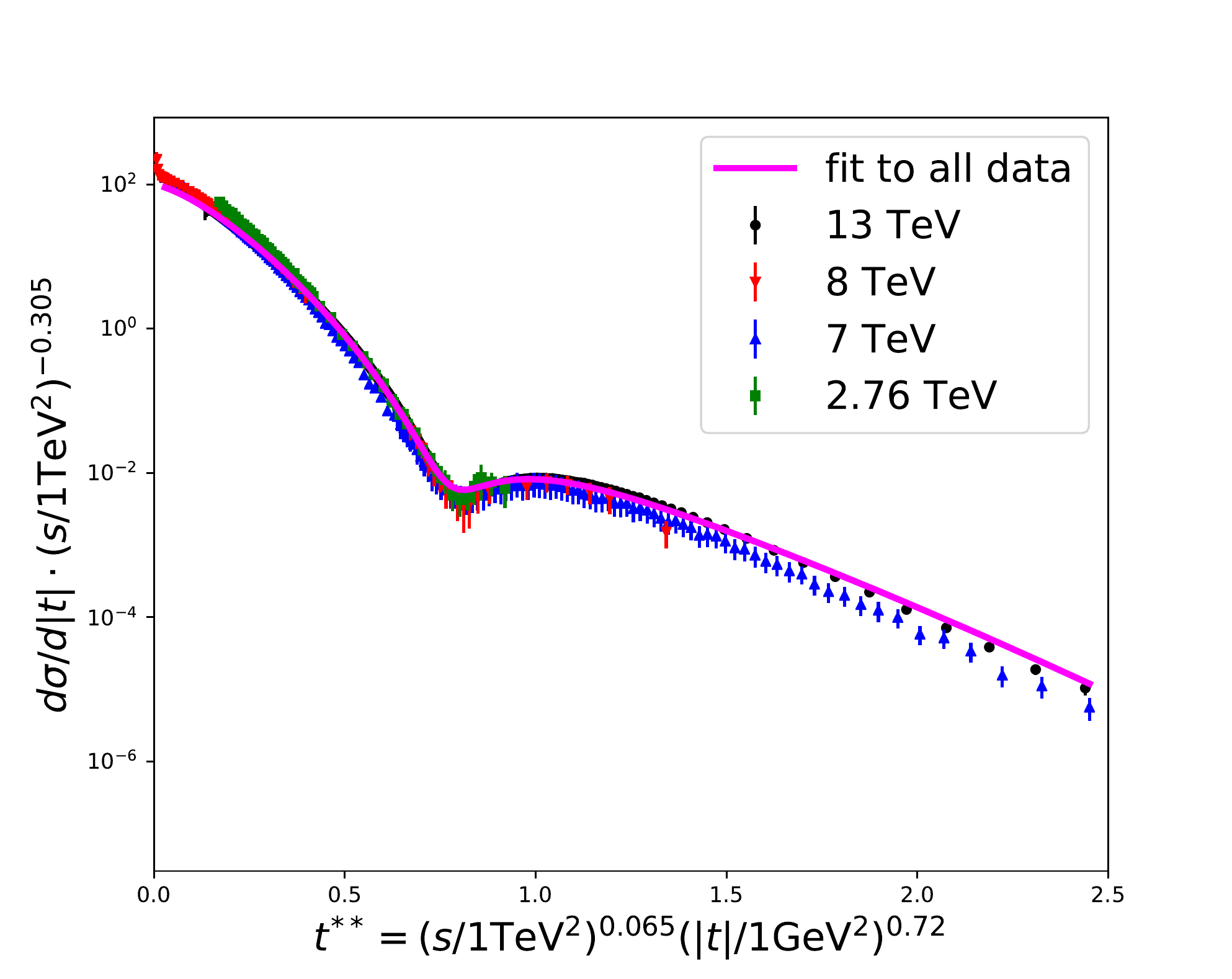}
\includegraphics[width=0.49\textwidth]{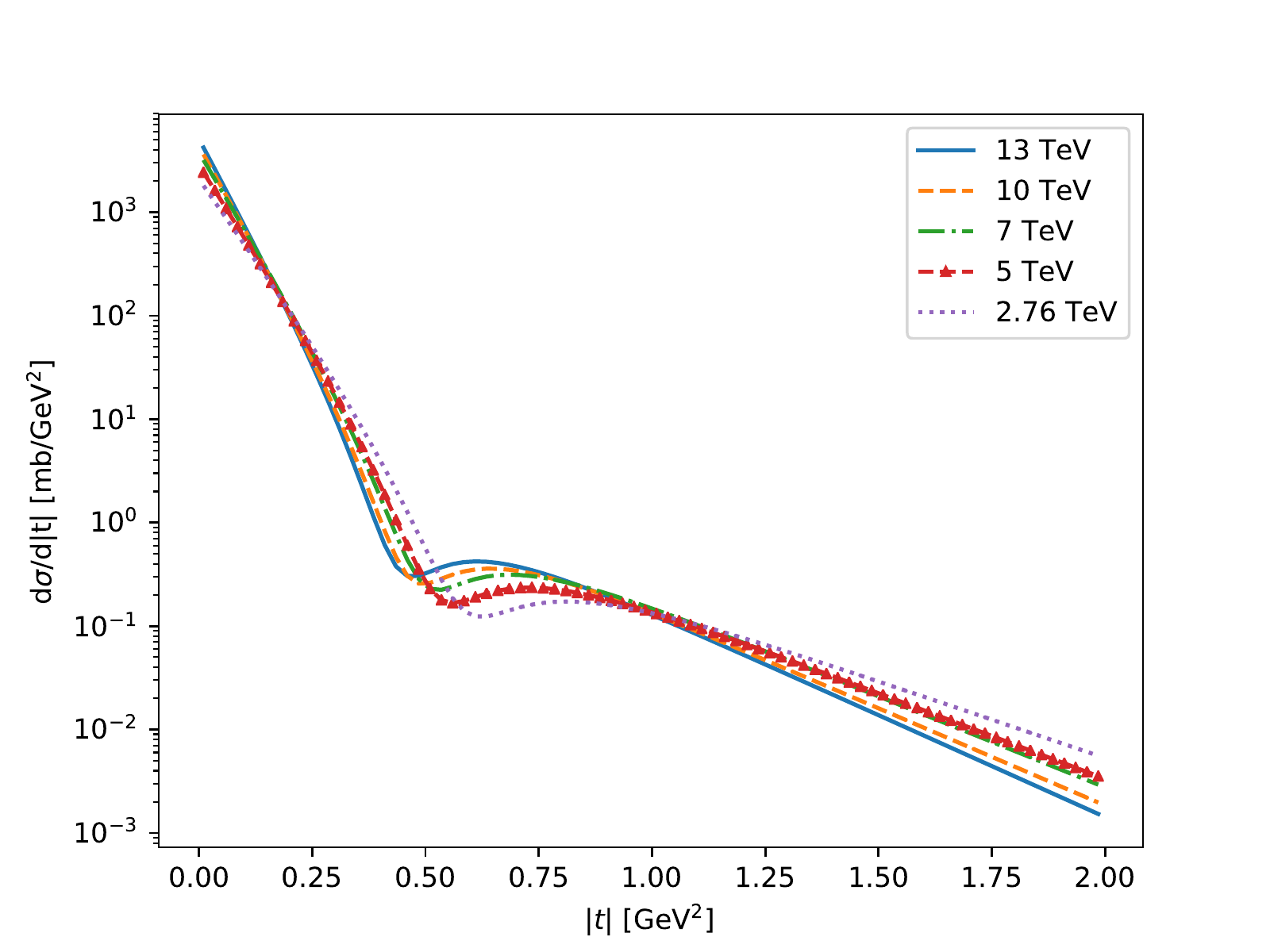}
\caption{Left: Fit to the scaled $ (s/$TeV$^2)^{-\alpha} d \sigma/d|t|$ from TOTEM elastic scattering data as a function of $t^{**}$. Right: Prediction of $\mathrm{d}\sigma/\mathrm{d}|t|$ from the scaling fit at different $\sqrt{s}$ } 
\label{sigmaAlphaVsTdoublestar}
\end{figure*}

\begin{figure*}
\centering
\includegraphics[width=0.47\textwidth]{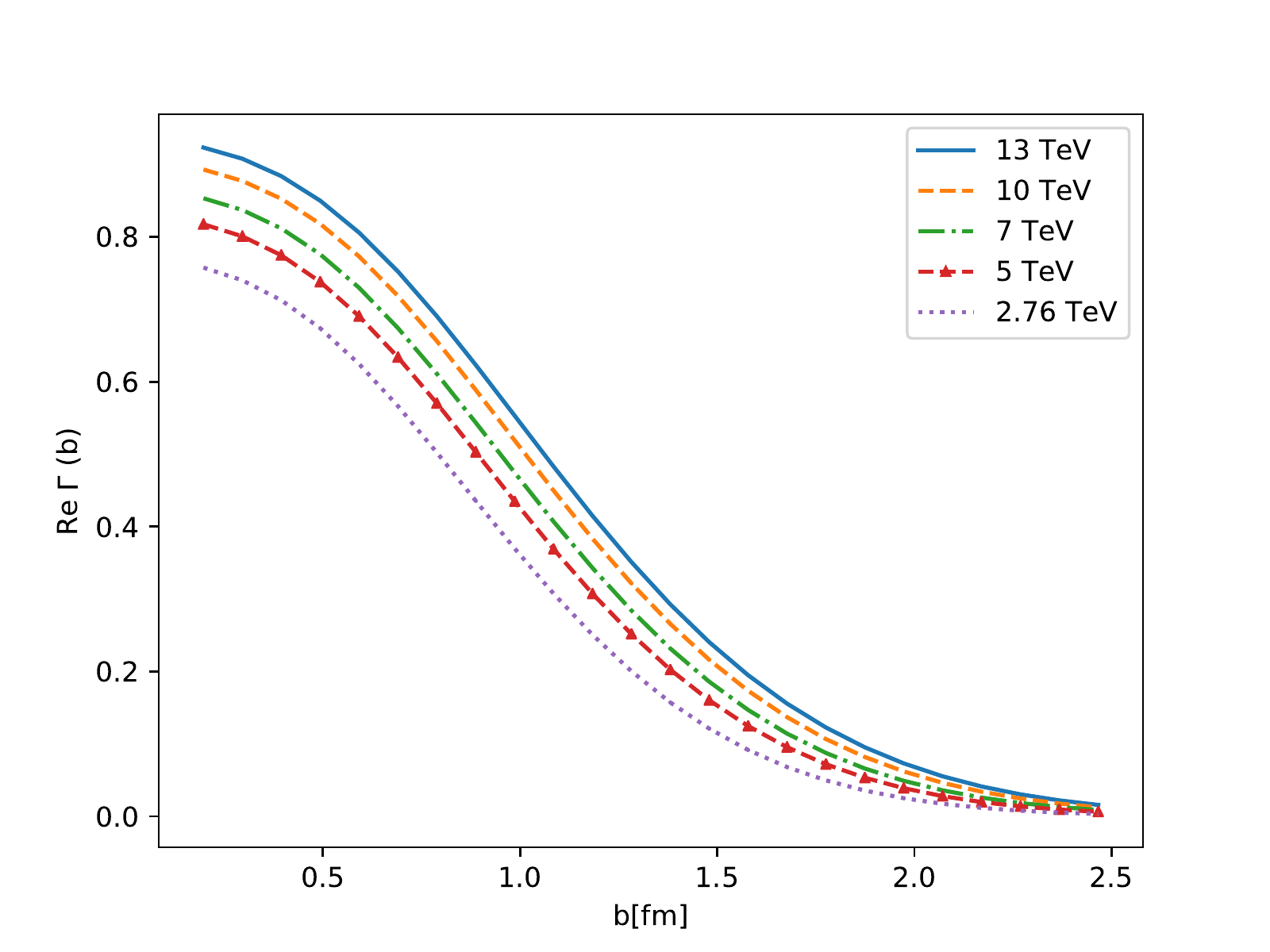}
\includegraphics[width=0.47\textwidth]{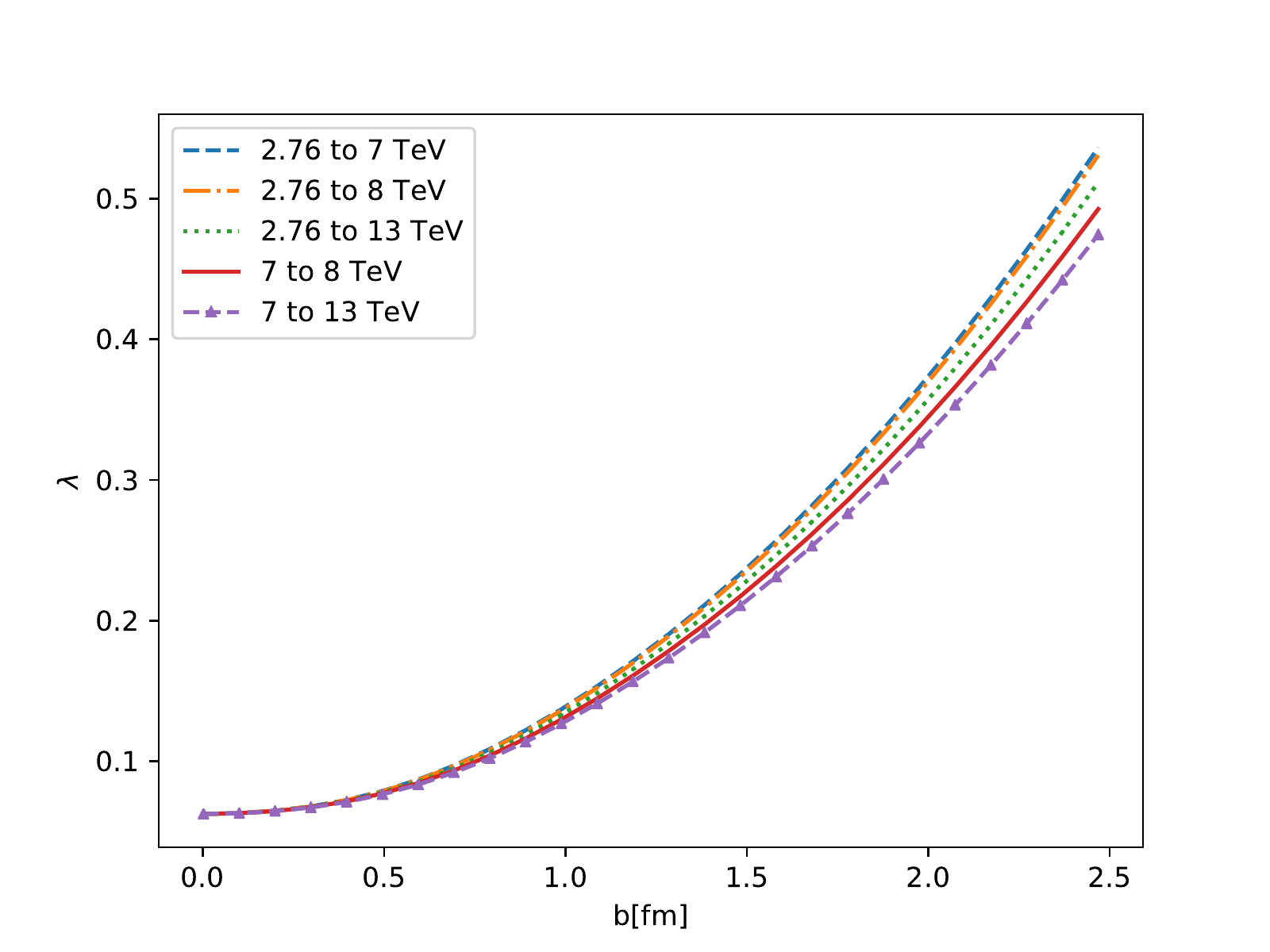} 
\caption{Left: The real part of the profile function $\mathrm{Re} \Gamma (b)$ as a function of the impact parameter ($b$) at different $\sqrt{s}$. Right: Power growth exponent $\lambda$ as a function of $b$ for various reference $\sqrt{s}$ pairs.}
\label{profile}
\end{figure*}

To facilitate calculations, we use a double exponential parametrization \cite{Phillips:1973dnz} to fit the elastic amplitude
\begin{equation}
\mathcal{A}(s,t) = i\big(\mathcal{A}_1 (s,t) + \mathcal{A}_2(s,t) \big)e^{i\theta}
\label{fitformula}
\end{equation}
where,
\begin{eqnarray}
\mathcal{A}_{1}(s,t) &=& N_1 (s) e^{-B_1(s)|t|} \\ \mathcal{A}_{2}(s,t) &=& N_2 (s) e^{-B_2(s)|t|}e^{i\phi} 
\end{eqnarray}

where $N_1(s) = N_1^0 (s/ \text{1 TeV}^2)^{\alpha/2}$, $N_2(s) = N_2^0 (s/ \text{1 TeV}^2)^{\alpha/2}$, $B_1(s) = B_1^0 (s/ \text{1 TeV}^2)^{\gamma/2}$ and $B_2(s) = B_2^0 (s/ \text{1 TeV}^2)^{\gamma/2}$, where $\alpha = 0.305 $ and $\gamma/2 \equiv  0.065/(1-A) = 0.065/0.72 \approx 0.09$ are the constants fixed by the scaling described in Section \ref{sec:section2}. The numerator and denominator of the definition of $\gamma/2$ correspond to the exponents of $s$ and $t$ in the definition of $t^{**} = (s/$1 TeV$^2)^{0.065} (|t|$/1 GeV$^2)^{0.72}$. The constant $\gamma$ is introduced to ease the fit parametrization. There are six free parameters: $N_1^0$, $N_2^0$, $B_1^0$, $B_2^0$, $\phi$, and $\theta$. The double exponential form allows for an analytical calculation of $\mathrm{Re}(\Gamma)$, as shown later in this section.

The fitting method is the following. We fit the data of the TOTEM Collaboration for all $\sqrt{s}$ in the scaling plane $\mathrm{d}\sigma/\mathrm{d}|t| (s/$TeV$^2)^{-0.305}$ as a function of $t^{**} = (s/$TeV$^2)^{0.065} (t/$GeV$^2)^{0.72}$ using the expressions in Eq.~\ref{sigmael} and \ref{fitformula}. We treat the experimental uncertainties as uncorrelated. The fit to all data in the scaling variables leads to $N_1^0 = 10.40 \pm 0.04$ $\sqrt{\mathrm{mb}}$ GeV$^{-1}$ , $N_2^0 = 0.49 \pm 0.01 $ $\sqrt{\mathrm{mb}}$ GeV$^{-1}$, $B_1^0 = 5.78 \pm 0.17 $ GeV$^{-2}$, $B_2^0 = 1.427 \pm 0.014 $ GeV$^{-2}$ and $\phi = -2.68 \pm 0.01$ rad. We fix $\theta = -0.09$ rad to satisfy $\rho(t =0 ) = \text{Re}(\mathcal{A})/\text{Im}(\mathcal{A}) = 0.10$ at high energies, as measured by the TOTEM experiment. The fit in the whole $t^{**}$ range yields $\chi^2/\text{ndof} = 8.7$ (599 data points). This tension is expected in this analysis, and is driven by the mismodeling of the low-$|t|$ region (there's no Coulomb-nuclear interference term in our fits) and the imperfections of scaling at low-$|t|$. In the dip and bump region, we have $\chi^2/\text{ndof} = 1.08$ for $0.2 < t^{**} < 1.5$ (476 data points). The TOTEM data together with the results of the fit are shown in Fig.~\ref{sigmaAlphaVsTdoublestar}, left. On Fig.~\ref{sigmaAlphaVsTdoublestar}, right, we display the values of $d \sigma/d|t|$ as predicted from the fit ($|t|$ dependence) and scaling ($s$ dependence).

We now turn to the calculation of $\text{Re}(\Gamma(s,b))$. The Fourier--Hankel transform of a gaussian is exact

\begin{equation}
F(b) = \int_0^{\infty} dq q J_0(qb) e^{-Bq^2}  = \frac{1}{2B} e^{\frac{-b^2}{4B}}     \; ,
\end{equation}

thus, the profile function is a sum of gaussians in $b$

\begin{eqnarray}
&~& \mathrm{Re} \Gamma(s,b) = \frac{1}{\hbar c \sqrt{\pi}} \Big(\frac{s}{1 \text{TeV}^2}\Big)^\frac{\alpha-\gamma}{2}  \times \Big( \cos(\theta) \frac{N_1^0}{2B_1^0} e^{\frac{-b^2}{4B_1^0 (s^{\gamma/2}s/\mathrm{TeV}^2)}}
+\cos(\theta+\phi) \frac{N_2^0}{2B_2^0} e^{\frac{-b^2}{4B_2^0 (s/\mathrm{TeV}^2)^{\gamma/2}}} \Big)\;,
\end{eqnarray}

We define the power energy exponent $\lambda$ to characterize the $s$-dependence of the profile function in impact parameter space as follows,

\begin{equation}
\frac{\mathrm{Re} \Gamma(s_1,b)}{\mathrm{Re} \Gamma(s_2,b)} = \bigg(\frac{s_1}{s_2}\bigg)^{\lambda}\; ,
\end{equation}
which leads to
\begin{equation}
 \lambda = \frac{1}{\ln (s_1/s_2)} \ln \Big(\frac{\mathrm{Re} \Gamma(s_1,b)}{\mathrm{Re} \Gamma(s_2,b)}\Big) \; .
\end{equation}

Here, $s_1$ and $s_2$ are two different squared center-of-mass energies (e.g., $\sqrt{s_1} = 13$ TeV and $\sqrt{s_2} = 7$ TeV). We refer to $s_1$ and $s_2$ as ``reference energies.'' The analytical expression for $\lambda$ can be computed from the known $\mathrm{Re} \Gamma$ at $s_1$ and $s_2$

\begin{eqnarray}
&~&    \lambda (b) = \frac{\alpha-\gamma}{2} +
    \frac{1}{\ln(s_1/s_2)}  \ln\Bigg( \frac{\cos(\theta)\frac{N_1^0}{B_1^0}e^{\frac{-b^2}{4B_1^0 (s_1/\mathrm{TeV}^2)^{\gamma/2}}}+\cos(\theta+\phi) \frac{N_2^0}{B_2^0} e^{\frac{-b^2}{4B_2^0 (s_1/\mathrm{TeV}^2)^{\gamma/2}}}}{\cos(\theta)\frac{N_1^0}{B_1^0}e^{\frac{-b^2}{4B_1^0 (s_2/\mathrm{TeV}^2)^{\gamma/2}}}+\cos(\theta+\phi) \frac{N_2^0}{B_2^0} e^{\frac{-b^2}{4B_2^0 (s_2/\mathrm{TeV}^2)^{\gamma/2}}}} \Bigg).
\end{eqnarray}
Scaling predicts a universal value of $\lambda$ at small impact parameter to be $\lambda (b \to 0) = \frac{\alpha-\gamma}{2} \approx 0.06$. In other words, the growth of the profile function with $s$ is independent of the reference energies $s_1$ and $s_2$ at low $b$. There is a weak dependence on the reference $\sqrt{s}$ used to calculate the energy power exponent, as shown in the right panel of Fig.~\ref{profile}.

A good approximation to $\lambda(b)$ can be derived by noting that $\cos(\theta) N_1^0/ B_1^0 \gg |\cos(\theta+\phi) N_2^0/ B_2^0| $ and that the first gaussian is shallower than the second one since $1/B_1^0 \ll 1/B_2^0$. After these considerations, the expression for $\lambda(b)$ simplifies to 
$$\lambda(b) \approx \frac{\alpha-\gamma}{2} + \frac{1}{4B_1^0\ln(s_1/s_2)} \Big(\frac{1}{(s_1/\text{TeV}^2)^{\gamma/2}}-\frac{1}{(s_2/\text{TeV}^2)^{\gamma/2}}\Big) b^2 .$$
$\lambda(b)$ is independent of the absolute normalization of the amplitudes and depends quadratically on $b$. The coefficient of $b^2$ evolves very slowly (quasi-logarithmically) with the reference energies $s_1$ and $s_2$.
The values of $\mathrm{Re}(\Gamma)$ and $\lambda$ as a function of the impact parameter $b$ are displayed in Fig.~\ref{profile}. We note some dependence of $\mathrm{Re}(\Gamma)$ with $s$ and the values of $\lambda$ as a function of the reference energies $s_1$ and $s_2$, as mentioned already from the analytical calculation.

It is instructive to analyze the total hadronic cross section $\sigma_{tot}$ predicted from the scaling behavior we found in $\mathrm{d}\sigma/\mathrm{d}|t|$, even if the agreement with the measured $\sigma_{tot}$ from TOTEM at 2.76, 7, 8 and 13 TeV is not expected since it is dominated by physics at low $|t|$, where scaling is not necessarily valid (we do not expect scaling to hold in the Coulomb region or in the region where Coulomb-nuclear interference effects are important). We find that $\sigma_\text{tot} = \frac{4\sqrt{\pi}}{\hbar c} (\cos(\theta)N_1^0+\cos(\theta+\phi)N_2^0) (s/\text{TeV}^2)^{\frac{\alpha}{2}} $
This calculation of $\sigma_\text{tot}$ overestimates the measurements reported by TOTEM by about a factor of $2$. This is again expected, since this could be a reflection of the mismodeling at low $|t|$ in our fits. The value of $\sigma_\text{tot}$ is directly related to the value of the elastic cross section at $t = 0$. Such an extrapolation requires proper treatment of the Coulomb interaction and Coulomb-nuclear interference effects. The scaling constant $\alpha$ drives the growth of the cross section with $s$, reminiscent of what is done with traditional Regge fits. We find the pomeron intercept for the total cross section from the scaling model to be $\alpha/2 \approx 0.15$. The information in $\sigma_\mathrm{tot}$ can be used as an additional constraint for future refinements of the scaling behavior we observe in $\mathrm{d}\sigma/\mathrm{d}|t|$.

\section{Summary} \label{sec:section4}

In this Letter, we analyzed the behavior of the differential cross section of proton-proton elastic scattering as a function of $t$ and $s$ at LHC energies. We found that the differential cross section for elastic scattering measured by the TOTEM Collaboration at $\sqrt{s} = $2.76, 7, 8, and 13 TeV exhibits scaling. The data fall onto a universal curve after mapping them with $ \mathrm{d}\sigma/\mathrm{d}t \to \mathrm{d}\sigma/\mathrm{d}t (s/\text{TeV}^2)^{-0.305}$ and $|t| \to (s/\text{TeV}^2)^{0.065} (|t|/\text{GeV}^2)^{0.72}$.
The quality of the scaling is very good in the entire range of $t^{**}$, except for very small and large values. When $t^{**}$ is very small (which corresponds to very small values of $|t|$), the elastic cross section is dominated by the quantum electrodynamics Coulomb and Coulomb-nuclear interference regions, and scaling (which is likely to be a property of quantum chromodynamics (QCD) ) is not expected to be valid in that region. When $t^{**}$ gets very large, we reach the perturbative QCD domain where saturation effects are expected to be weaker.

We also explored the implications of this approximate scaling law in the impact parameter picture of the scattering amplitudes. We have extracted the exponent $\lambda$ that governs the energy dependence of the profile function as a function of impact parameter $b$. We found that this exponent has values of $\lambda = 0.06$ at small values of $b$ and up to $\lambda = 0.4$ at large values of $b = 2$ fm. The power exponent $\lambda$ has a weak dependence on the reference energies $\sqrt{s}$ used to extract it. In particular, at $b = 0$, the value of $\lambda = 0.06$ does not depend on the reference energies and is directly related to the scaling properties.

It is also worth noting that the $\lambda$ values at small $b$ are compatible with expectations from a dense object, such as a black disc, see for example  \cite{Frankfurt:2004fm}, and reach higher values around 0.3 for $b = 1$ fm. The latter is reminiscent of the power-law exponent in the small-$x$ limit of QCD, described by the perturbative Balitsky--Fadin--Kuraev--Lipatov (BFKL) evolution equation at next-to-leading logarithmic accuracy ~\cite{Balitsky:1978ic,Kuraev:1977fs,Lipatov:1985uk}, see also \cite{Avsar:2007ht,Flensburg:2008ag}.

To conclude, let us try to give a possible qualitative physics interpretation of the scaling that we observe, which could be explored in future studies. Elastic hadron-hadron interactions are mediated by colorless exchanges, dominated by pomeron and odderon exchange at high energies. One could assume that these are due to the existence of colorless gluonic compounds in the proton, whose density in the proton may vary slowly as a function of $s$. Such energy behavior could be related to that of elastic cross section with $s$ \cite{Donnachie:2013xia} ($t^{**} $ scales with $s^{0.065}$). In this picture, the scaling found in this Letter together with the value of $\lambda$ at low $b$, could be interpreted as having a large density of gluons inside these gluonic compounds that reach the black disc limit at small $b$. At higher $b$, the density of gluons in the ``hot spot'' is smaller and in principle describable by BFKL dynamics. In this sense, we can interpret our results as the presence of hot spots~\cite{Bartels:1991tf} in the proton at high energy. The density of these hot spots in the proton can be small, but the density of the gluons inside these hot spots can be large. In order to probe this idea further, we would have to implement it in a full model description, which goes beyond the scaling observation presented in this Letter. Previous implementations of hot spots models with spatial correlations to understand elastic scattering, particularly in the context of the ``hollowness'' effect, have been presented in Refs.~\cite{Albacete:2016pmp, Albacete:2017ajt}.

Generally speaking, the scaling behavior we observe might be also indicative of underlying general properties of high-energy scattering in the context of Regge theory when the leading trajectory exchanges are pomeron and odderon trajectories.

\section*{Acknowledgements}

We thank Balt van Rees, Cyrille Marquet, Martin Hentschinski, and Mark Strikman for useful discussions.

\bibliographystyle{ieeetr}
\bibliography{plb.bib}

\end{document}